\documentclass[11pt,a4paper,final]{iopart}

\usepackage[T1]{fontenc}
\usepackage[utf8]{inputenc}
\usepackage[english]{babel}
\usepackage{iopams}
\usepackage{bbm}
\usepackage{graphicx}
\usepackage[colorlinks,
           linkcolor=blue,
           citecolor=blue,
           urlcolor=blue,
           bookmarks=true,        
           bookmarksopen=true,    
           bookmarksnumbered=true,
]{hyperref}
\usepackage{xcolor}

\usepackage{etoolbox}
\makeatletter
\newcommand{\mainmatter}{%
  \setcounter{footnote}{0}%
  \patchcmd{\@makefntext}{\fnsymbol}{\arabic}{}{}%
  \patchcmd{\@thefnmark}{\fnsymbol}{\arabic}{}{}%
  \def\@makefnmark{\textsuperscript{\arabic{footnote}}}%
}
\makeatother

\global\long\def\ket#1{\left| #1\right\rangle }
\global\long\def\bra#1{\left\langle #1 \right|}
\global\long\def\kket#1{\ket{\ket{#1}}}
\global\long\def\bbra#1{\bra{\bra{#1}}}
\global\long\def\Tr{\mathrm{Tr}}
\global\long\def\pd{\partial}
\global\long\def\im{\mathrm{Im}}
\global\long\def\re{\mathrm{Re}}
\global\long\def\abs#1{\left|#1\right|}
\renewcommand*\d{\mathop{}\!\mathrm{d}}

\begin{document}

\title{Spectral and Steady-State Properties of Random Liouvillians}

\author{Lucas  S\'a$^1$, Pedro Ribeiro$^{1,2}$, and  Toma\v z Prosen$^3$}

\address{
$^1$CeFEMA, Instituto Superior T\'ecnico, Universidade de Lisboa, Av.\ Rovisco Pais, 1049-001 Lisboa, Portugal\\
$^2$Beijing Computational Science Research Center, Beijing 100193, China\\
$^3$Department of Physics, Faculty of Mathematics and Physics, University of Ljubljana, Ljubljana, Slovenia}

\ead{lucas.seara.sa@tecnico.ulisboa.pt, ribeiro.pedro@tecnico.ulisboa.pt, and tomaz.prosen@fmf.uni-lj.si}

\begin{abstract}
We study generic open quantum systems with Markovian dissipation, focusing on a class of stochastic Liouvillian operators of Lindblad form with independent random dissipation channels (jump operators) and a random Hamiltonian.
We establish that the global spectral features, the spectral gap, and the steady-state properties follow three different regimes as a function of the dissipation strength, whose boundaries depend on the particular quantity. 
Within each regime, we determine the scaling exponents with the dissipation strength and system size. 
We find that, for two or more dissipation channels, the spectral gap increases with the system size. 
The spectral distribution of the steady state is Poissonian at low dissipation strength and conforms to that of a random matrix once the dissipation is sufficiently strong. 
Our results can help to understand the long-time dynamics and steady-state properties of generic dissipative systems.
\end{abstract}

\mainmatter

\section{Introduction}

Although open quantum systems are strongly influenced by their environments, a description of all environment degrees of freedom is often impossible and unnecessary. 
Master equations describing the evolution of the system's reduced density matrix, $\pd_{t}\rho=\mathcal{L\left(\rho\right)}$, where the Liouvillian acquires the Lindblad form~\cite{Breuer2002}
\begin{equation}\label{eq:Lindblad}
\mathcal{L} =\mathcal{L}_{H}+\sum_{\ell}\mathcal{D}_{W_{\ell}},
\end{equation}
provide a simple approach to model open quantum dynamics. Here, $\mathcal{L}_{H}\left(\rho\right)=-i\left[H,\rho\right]$ represents the unitary evolution under the Hamiltonian $H$, and 
\begin{equation}
    \mathcal{D}_{W_{\ell}}\left(\rho\right)=W_{\ell}\rho W_{\ell}^{\dagger}-\frac{1}{2}W_{\ell}^{\dagger}W_{\ell}\rho-\frac{1}{2}\rho W_{\ell}^{\dagger}W_{\ell}
\end{equation}
the contribution of each dissipation channel by the action of the jump operator $W_{\ell}$. 

This approach assumes either a weakly coupled or strongly coupled environment~\cite{Breuer2002}, with memory times much shorter than all other characteristic energy scales. Due to this so-called Markovian assumption, Lindblad dynamics fails to capture certain processes, such as the ones responsible for coherent low-temperature transport~\cite{Ribeiro2014e}. Nevertheless, the method is widely used in areas ranging from thermodynamics of quantum engines~\cite{Hofer2017} to the description of quark-gluon plasma~\cite{Boni2017}. Perhaps, its most important application is to model quantum optic setups, which typically fulfill the required conditions as driving fields shift the relevant energies of the environment to a region where the spectral density is large~\cite{GardinerZoller}.

Although the Lindblad form significantly simplifies the problem, obtaining dynamic and steady-state properties given a Lindblad operator remains a major theoretical challenge. In one dimension, efficient numerical approaches have been developed based on matrix product operator ideas~\cite{Verstraete2004,Prosen2009}. An alternative strategy, that has recently been extremely successful, consists of studying exactly solvable (or integrable) models~\cite{Prosen2008a,Prosen2011a,Prosen2014,Medvedyeva2016,Rowlands2017a,Ribeiro2019}.
Albeit enlightening, integrable Lindbladians, as their Hamiltonian counterparts, are expected to have very peculiar properties and remain a set of measure zero among all possible Lindbladian dynamics.  For generic cases, exact diagonalisation of relatively small systems remains the only option. 

Tools to study generic Hamiltonians are available for non-integrable closed quantum systems. They rely on the widely supported conjecture~\cite{Bohigas1984} that universal features of spectral and eigenstate properties of quantum systems with a well-defined chaotic classical limit follow those of random matrix theory (RMT)~\cite{wigner1955,beenaker1997,guhr1998,haake2013,stockmann2000,gomez2011}. This is to be contrasted with integrable systems where level spacings typically follow Poisson statistics~\cite{berry1977}. RMT has been reported to also describe complex many-body systems without well defined classical correspondents~\cite{alessio2016,kos2018,chan2018}. These predictions are particularly appealing as they are insensitive to the microscopic details of particular models, relying solely on symmetry properties of the Hamiltonian, i.e.\ to which: Gaussian unitary (GUE), Gaussian orthogonal (GOE), or Gaussian symplectic (GSE) ensemble, it belongs to. In view of the tremendous success of RMT, it is natural to ask if a similar approach can be followed in the case of Lindbladian dynamics. While open chaotic scattering~\cite{sommers1998,haake1992,lehmann1995} and quantum dissipation and decoherence~\cite{lutz1999,gorin2003,gorin2008,moreno2015,pineda2015,xu2019} have been studied using RMT in the past, this route has remained essentially unexplored up to very recently~\cite{denisov2018,can2019a,can2019}. 

Following these general ideas, ~\cite{denisov2018} studies an ensemble of random Lindblad operators with a maximal number of independent decay channels and finds that, in this case, the spectrum acquires a universal lemon-shaped form. In a similar spirit,~\cite{can2019a} considers an ensemble of random Lindbladians consisting of a Hamiltonian part and a finite number of Hermitian jump operators and finds a sharp spectral transition as a function of the dissipation strength. Finally,~\cite{can2019} studies in detail the spectral gap for several different Liouvillians. Yet, there is a number of open questions related to the nature of the spectrum, specifically when the dissipative and the Hamiltonian components are comparable. Additionally, when the jump operators are Hermitian, the steady state is the infinite-temperature thermal state (i.e.\ proportional to the identity operator). The properties of the non-trivial steady state, ensuing in the presence of non-Hermitian jump operators, are completely unexplored.

The aim of this paper is to study the universal properties of the spectrum and the steady state of Liouvillian operators consisting of a Hamiltonian component and a set of dissipation channels, taken from appropriate ensembles of random matrices. We consider a finite number of non-Hermitian jump operators allowing for a non-trivial steady state. As a function of the effective dissipation strength, $g_{\mathrm{eff}}$, we find a rich set of regimes regarding global spectral features such as the spreading of the decay rates~$X$, the spectral gap~$\Delta$, and the spectral properties of the steady-state density matrix~$\rho_{0}$. Each regime is characterised by a scaling of the corresponding quantity with the system size~$N$ and with $g_{\mathrm{eff}}$. The finite-size scaling of the boundaries between different regimes is also determined. 

We expect these results to model a very broad class of nonregular Markovian open quantum systems. Indeed, in view of the quantum chaos conjecture, systems effectively modeled by random Lindbladians should be the rule rather than the exception. The predictive power of the present analysis relies on the fact that, once a regime is determined, the system's universal properties can be obtained solely by symmetry arguments.

The paper is organised as follows. First, we describe an unbiased construction of a random Liouvillian in Section~\ref{sec:Liouv_construction}. Second, we numerically study its spectral (Section~\ref{sec:spectrum}) and steady-state (Section~\ref{sec:steady_state}) properties. Finally, we end with a short summary of our findings and their implications for determining the properties of generic Markovian dissipative systems in Section~\ref{sec:conclusion}. Several appendices contain details of analytical derivations and additional numerical results.

\section{Random Liouvillian Ensembles}
\label{sec:Liouv_construction}

To obtain a suitable set of random Liouvillians of the Lindblad form, we define a complete orthogonal basis, $\left\{ G_{i}\right\} $ with $i=0,\dots,N^{2}-1$, for the space of operators acting on an Hilbert space of dimension $N$\footnote{
We consider only systems with finite-dimensional local Hilbert spaces. Below, we are interested in taking the thermodynamic limit, which is to be understood as $N\to\infty$.}, 
respecting $\Tr\left[G_{i}^{\dagger}G_{j}\right]=\delta_{ij}$, with $G_{0}=\mathbbm{1}/\sqrt{N}$ proportional to the identity. Each jump operator can be decomposed as $W_{\ell}=g\sum_{j=1}^{N^{2}-1}G_{j}w_{j\ell}$. Note that $W_{\ell}$ is taken to be traceless, i.e.\ orthogonal to $G_{0}$, to ensure that the dissipative term in (\ref{eq:Lindblad}) does not contribute to the Hamiltonian dynamics. In the $\left\{ G_{i}\right\}$ basis, the Liouvillian is completely determined by matrices $H$ and $w$, 
\begin{equation}
\label{eq:lindblad-1}
\fl\qquad
\mathcal{L}\left(\rho\right)=-i\left[H,\rho\right]
+g^{2}\sum_{j,k=1}^{N^{2}-1}d_{jk}\left\{ G_{j}\rho G_{k}^{\dagger}-\frac{1}{2}\left[\rho G_{k}^{\dagger}G_{j}+G_{k}^{\dagger}G_{j}\rho\right]\right\},
\end{equation}
where $d_{jk}=\sum_{\ell=1}^{r}w_{j\ell}w_{k\ell}^{*}=(ww^{\dagger})_{jk}$ is an $(N^2-1)\times(N^2-1)$ positive-definite matrix. We denote by $r$ the number of jump operators in (\ref{eq:Lindblad}) (i.e.\ $\ell=1,\dots,r$) which counts the number of independent system operators coupled to independent environment degrees of freedom. To obtain a random Liouvillian, we draw $H$ from a Gaussian ensemble with unit variance~\cite{mehta2004,haake2013}, i.e.\
\begin{equation}
P_{N}\left(H\right)\propto \exp\left\{-\frac{1}{2}\Tr\left(H^{2}\right)\right\},   
\end{equation} 
and $w$ from a Ginibre ensemble~\cite{ginibre1965,haake2013}\footnote{For each realisation of the system, our operators are time-independent. The statistical approach then corresponds to ensemble-averaging over random matrices. This is to be contrasted with the different approach of time-averaging perturbations that evolve randomly in time, but have a fixed direction in matrix-space.}, i.e.\
\begin{equation}
P_{\left(N^{2}-1,r\right)}\left(w\right)\propto \exp\left\{-\frac{1}{2}\Tr\left(w^{\dagger}w\right)\right\}. 
\end{equation}
The coupling constant $g>0$ parameterizes the dissipation strength (when the typical scale of the frequencies of $H$ is of order one, then the typical scale of the decay rate of $\mathcal{D}_{W_\ell}$ is of order $g^2$). We consider two cases: real matrices, $H_{ij}=H_{ji},w_{ij}\in\mathbb{R}$ (labeled by $\beta=1$ in the following), and complex matrices, $H_{ij}=H_{ji}^{*},w_{ij}\in\mathbb{C}$ ($\beta=2$). 

We study the statistical properties of the spectrum of $\mathcal{L}$ drawn from an ensemble of random Lindblad operators parameterised by $N$, $r$, $\beta$ and $g$. The right eigenvectors of $\mathcal{L}$, respecting $\mathcal{L}\left(\rho_{\alpha}\right)=\Lambda_{\alpha}\rho_{\alpha}$, with $\alpha=0,\dots,N^{2}-1$, are denoted by $\rho_{\alpha}$, with $\Lambda_{\alpha}$ the respective eigenvalue.
By construction, $\re\left(\Lambda_{\alpha}\right)\le\Lambda_{0}=0$, and $\rho_{0}$, if unique\footnote{The steady state is unique in the absence of any additional symmetries. Since we are considering the less structured Liouvillian possible the steady states we find are unique by construction.}, is the asymptotic steady state, left invariant by the evolution. If an experiment probes a dissipative system for long-enough timescales, it is this state it studies.

In Figure~\ref{fig:spectrum}~$(b)$, we represent several relevant energy- (or inverse time-) scales. 
The spectral gap, $\Delta=\min_{\alpha>0}\re\left(-\Lambda_{\alpha}\right)$ describes the typical time it takes the system to reach the steady state, i.e.\ the duration of transient effects.
The variance along the real axis, $X^{2}=\sum_{\alpha}\left[\re\left(\Lambda_{\alpha}-R\right)\right]^{2}/N^{2}$, sets the spread of decay rates and also the typical minimum time to observe the onset of dissipation and decoherence. 
The variance along the imaginary axis, $Y^{2}=\sum_{\alpha}\left(\im\Lambda_{\alpha}\right)^{2}/N^{2}$, gives the timescale for the oscillations of the states' phases.
We also depict the center of mass of the spectrum, $R=\sum_{\alpha}\Lambda_{\alpha}/N^{2}$, which can usually be trivially shifted away.

The spectrum and eigenvectors of $\mathcal{L}$ are obtained by exact diagonalisation.

\section{Spectral properties}
\label{sec:spectrum}

\begin{figure}[tbp]
\centering 
\includegraphics[width=\textwidth]{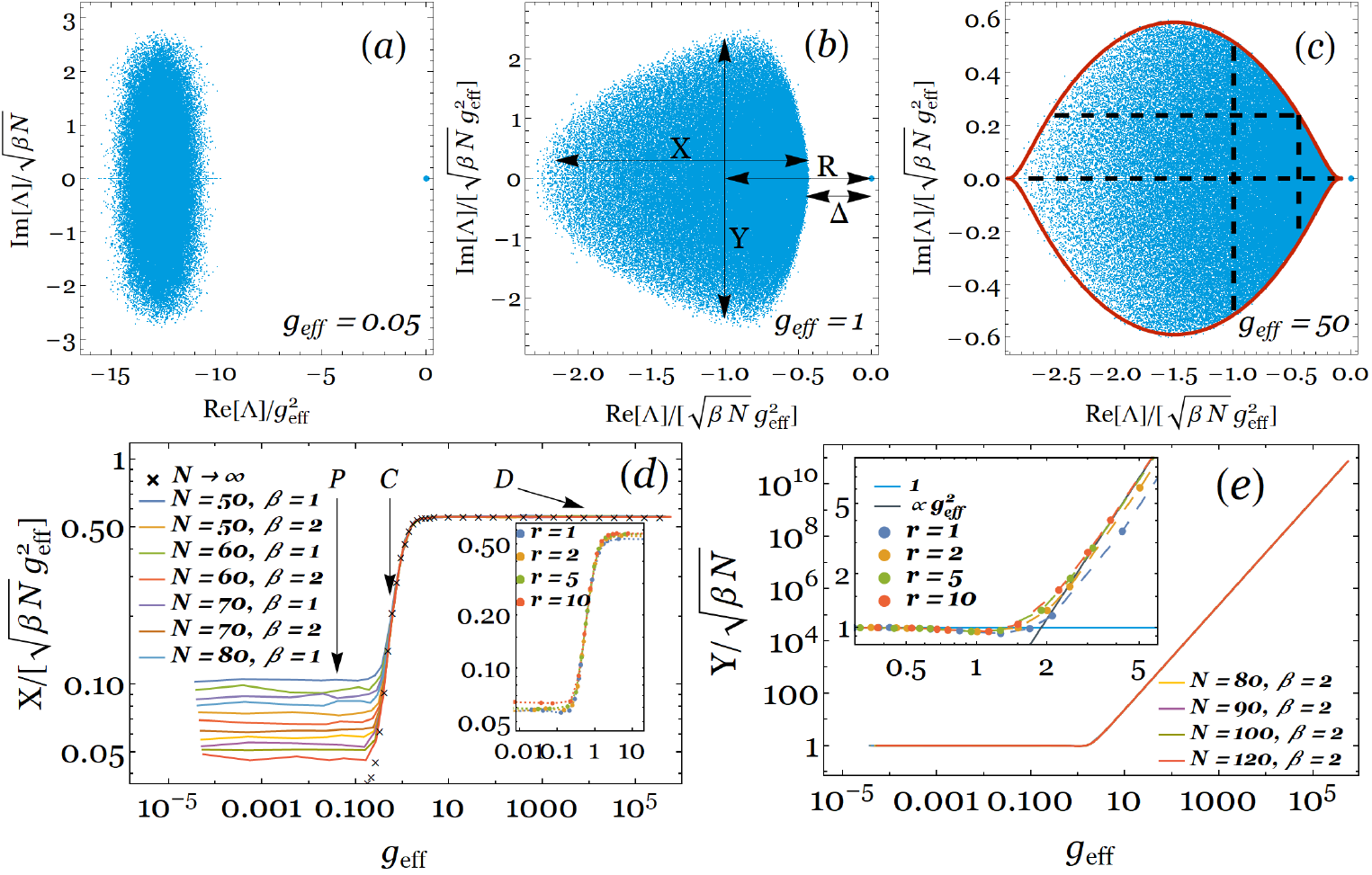}
\caption{\label{fig:spectrum} $(a)-(c)$ Spectrum of a random Lindblad operator for different values of $g_\mathrm{eff}$, computed for $N=80$, $\beta=2$ and $r=2$. $R$, $X$, $Y$ and $\Delta$ are, respectively, the center for mass of the spectrum, the standard deviation along the real and imaginary axes, and the spectral gap. The horizontal (vertical) dashed lines in $(c)$ correspond to $0$, $Y$ ($R$, $R+X$), and the solid boundary is explicitly computed in~\cite{denisov2018}. $(d)$ and $(e)$ show the scaling of $X$ and $Y$ for different values of $N$ and $r=2$ as a function of $g_{\mathrm{eff}}=\left(2r\beta N\right)^{1/4}g$. The insets of $(d)$ and $(e)$ show the same information, but for different values of $r$ and fixed $N=80$.}
\end{figure}

Figures~\ref{fig:spectrum} $(a)$--$(c)$ show the spectrum of a random Lindblad operator in the complex plane computed for different values of $g$. 
The boundaries of the spectrum evolve from an ellipse, for small $g$, to a lemon-like shape at large $g$. 
In Figure~\ref{fig:spectrum}~$(c)$ we plot (solid line) the spectral boundary, explicitly computed in~\cite{denisov2018} for the case $r=N^{2}-1$\footnote{
The spectral boundary from~\cite{denisov2018} is centered at the origin and has to be displaced by a shift $-(1+r)/(2\sqrt{r})$ and then rescaled by $\sqrt{2}$ to give the line of Figure~\ref{fig:spectrum}~$(c)$.}.
These results, obtained here for $r=2$, indicate that the lemon-shaped spectral boundary is ubiquitous in the strong dissipation regime.

\begin{figure}
\centering \includegraphics[width=\textwidth]{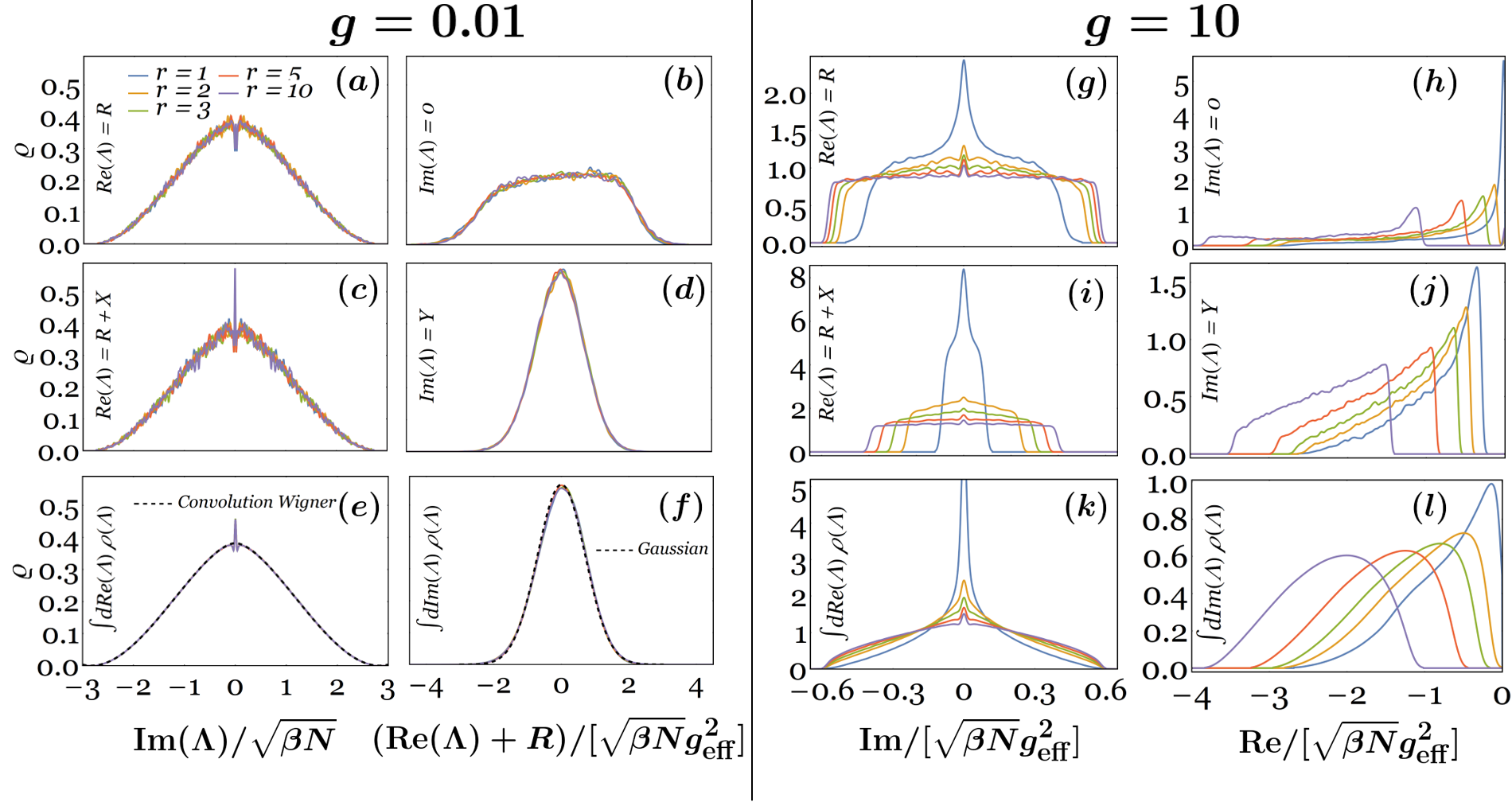}
\caption{\label{fig:spectrum_cuts} Spectral density along the cuts $\re\left(\Lambda\right)=R$, $R+X$ and $\im\left(\Lambda\right)=0$, $Y$, computed for $N=80$, $\beta=2$, several values of $r$ and two values of $g$, such that the system is in regime P: $(a)-(d)$; and in regime D: $(g)-(j)$. $(e)$, $(f)$: integrated density along the real and imaginary axes, $\varrho_{\mathrm{I}}\left(\Lambda_{\mathrm{I}}\right)$ and $\varrho_{\mathrm{R}}\left(\Lambda_{\mathrm{R}}\right)$, for regime P. In $(b)$, $(d)$ and $(f)$ the zero eigenvalue is omitted. $(k)$ and $(l)$: same quantities for regime D.}
\end{figure}

Figures~\ref{fig:spectrum} $(d)$ and $(e)$ show the scaling of $X$
and $Y$ with $N$ as a function of  
\begin{equation}
g_{\mathrm{eff}}=\left(2r\beta N\right)^{1/4}g.
\end{equation}
The three representative points (P, C, D) of Figure~\ref{fig:spectrum} $(d)$ correspond to regimes for which $X$ depicts a qualitatively different behaviour. The perturbative regime P corresponds to the weak-coupling limit of the Lindblad equation, while the dissipative regime D relates to the singular-coupling limit.
The observed scaling collapse shows that, for large $g_{\mathrm{eff}}$, the standard deviation along the real axis behaves as 
\begin{equation}
    X\propto\left[(\beta N)^{1/4}g_{\mathrm{eff}}\right]^{2}f_{X}^{>}(g_{\mathrm{eff}}),
\end{equation} 
with $f_{X}^{>}$ an unknown scaling function satisfying $f_{X}^{>}(x)\propto x^{0}$ for $x\!\to\!\infty$ and $f_{X}^{>}(x)\!\propto\!x^{2}$ for $x\to0$. A similar scaling collapse can be obtained for small $g_{\mathrm{eff}}$ (not shown, see~\ref{app:spectrum}) yielding 
\begin{equation}
    X\simeq g_{\mathrm{eff}}^{2}f_{X}^{<}\left[(\beta N)^{1/4}g_{\mathrm{eff}}\right],
\end{equation}
for which the scaling function satisfies $f_{X}^{<}(x)\propto x^{0}$ when $x\!\to\!0$ and $f_{X}^{<}(x)\propto x^{2}$ when $x\!\to\!\infty$. The asymptotic power law behaviour of the scaling functions is hard to determine for the available values of $N\leq120$, however, it is fully compatible with the large-$N$ extrapolation both in the large- and small-$g_{\mathrm{eff}}$ regimes (see black crosses in Figure~\ref{fig:spectrum}~$(d)$ and also~\ref{app:spectrum}). This result is further corroborated by the compatibility conditions (for further details see \ref{app:exponents})
\begin{equation}
    \lim_{g_{\mathrm{eff}}\to0}(\beta N)^{1/2}f_{X}^{>}(g_{\mathrm{eff}})\simeq\lim_{(\beta N)^{1/4}g_{\mathrm{eff}}\to\infty}f_{X}^{<}\left[(\beta N)^{1/4}g_{\mathrm{eff}}\right].
\end{equation}
We can thus identify the three regimes: P, for $g_{\mathrm{eff}}\lesssim\left(\beta N\right){}^{-1/4}$; C, for $(\beta N)^{-1/4}\lesssim g_{\mathrm{eff}}\lesssim(\beta N)^{0}$; and D, for $(\beta N)^{0}\lesssim g_{\mathrm{eff}}$, corresponding to each representative point. Note that the rescaled quantities plotted in Figure~\ref{fig:spectrum} $(d)$ and $(e)$ do not seem to depend on the index $\beta$. There is a small $r$-dependence, which converges rapidly for increasing $r$ (see insets). For the standard deviation along the imaginary axes we find, $Y\simeq\sqrt{\beta N}$ for regimes P and C, and $Y\propto\sqrt{\beta N}g_{\mathrm{eff}}^{2}$ within regime D. Therefore, the variance in regime P can be explained by a perturbative treatment of the dissipative term---the value of $Y$ corresponds to the unit variance of the random Hamiltonian, and $X\propto g_{\mathrm{eff}}^{2}$ is expected from a degenerate perturbation theory.

Figures~\ref{fig:spectrum_cuts} $(a)$--$(f)$ and $(g)$--$(l)$ show the spectral density, $\varrho\left(\Lambda\right)=\sum_{\alpha}\delta^{2}\left(\Lambda-\Lambda_{\alpha}\right)$, along the cuts depicted in Figure~\ref{fig:spectrum} $(c)$ for regimes P and D, respectively, for various numbers of jump operators. Results for $\beta=1$ show the same limiting behaviour in the large-$N$ limit, after proper rescaling. For P, the density along $\re\left(\Lambda\right)=R$ and $R+X$, does not depend on the number of jump operators, $r$. The integrated distribution of the imaginary parts, $\varrho_{\mathrm{I}}\left(\Lambda_{\mathrm{I}}\right)=\int\d^{2}\Lambda\,\delta\!\left(\Lambda_{\mathrm{I}}-\im\Lambda\right)\varrho\left(\Lambda\right)$, is well described by a convolution of Wigner's semicircle laws, $\varrho_{\mathrm{W}}(E)\simeq \sqrt{E_{*}^2-E^2}$, i.e.\ 
\begin{equation}
\varrho_{\mathrm{I}}\left(\Lambda_{I}\right)\simeq\int\d E_{1}\d E_{2}\,\delta\!\left(\Lambda_{I}-E_{1}+E_{2}\right)\varrho_{\mathrm{W}}\left(E_{1}\right)\varrho_{\mathrm{W}}\left(E_{2}\right),    
\end{equation}
except at $\Lambda_\mathrm{I}=0$ where there is an increase of spectral weight, which is depleted from the immediate vicinity of the real axis. This last feature holds for general hermiticity-preserving operators, which can be brought to a real representation by a trivial similarity transformation. They are therefore related to the real Ginibre ensemble, where it can be explicitly shown~\cite{forrester2007}. The spectral weights along the cuts $\im\left(\Lambda\right)=0$ and $Y$ scale with $\sqrt{r}$ and the integrated distribution of the real parts, $\varrho_{\mathrm{R}}\left(\Lambda_{\mathrm{R}}\right)=\int\d^{2}\Lambda\,\delta\!\left(\Lambda_{\mathrm{R}}-\re\Lambda\right)\varrho\left(\Lambda\right)$, is well approximated by a Gaussian. As for the variance, in the P regime, these results can be derived from a perturbative small-$g$ expansion. In the strongly dissipative regime, D, the spectral density along $\re\left(\Lambda\right)=R$ and $R+X$ depends on $r$ but converges rapidly to the $r\to\infty$ limit. 

The case $r=1$ is qualitatively different from $r>1$. This can also be observed in the $\im\left(\Lambda\right)=0$ and $Y$ cuts, and in $\varrho_{\mathrm{R}}$, where for $r=1$ the spectral weight is finite for $\Lambda_{\mathrm{R}}\to0^{-}$, in contrast with the $r>1$ results that develop a spectral gap. We also found different scaling properties for $r=1$ compared to $r>1$, see below. Presently, the physical reason behind this discrepancy between $r=1$ and $r>1$ is unknown.

\begin{figure}[tbp]
\centering \includegraphics[width=0.65\textwidth]{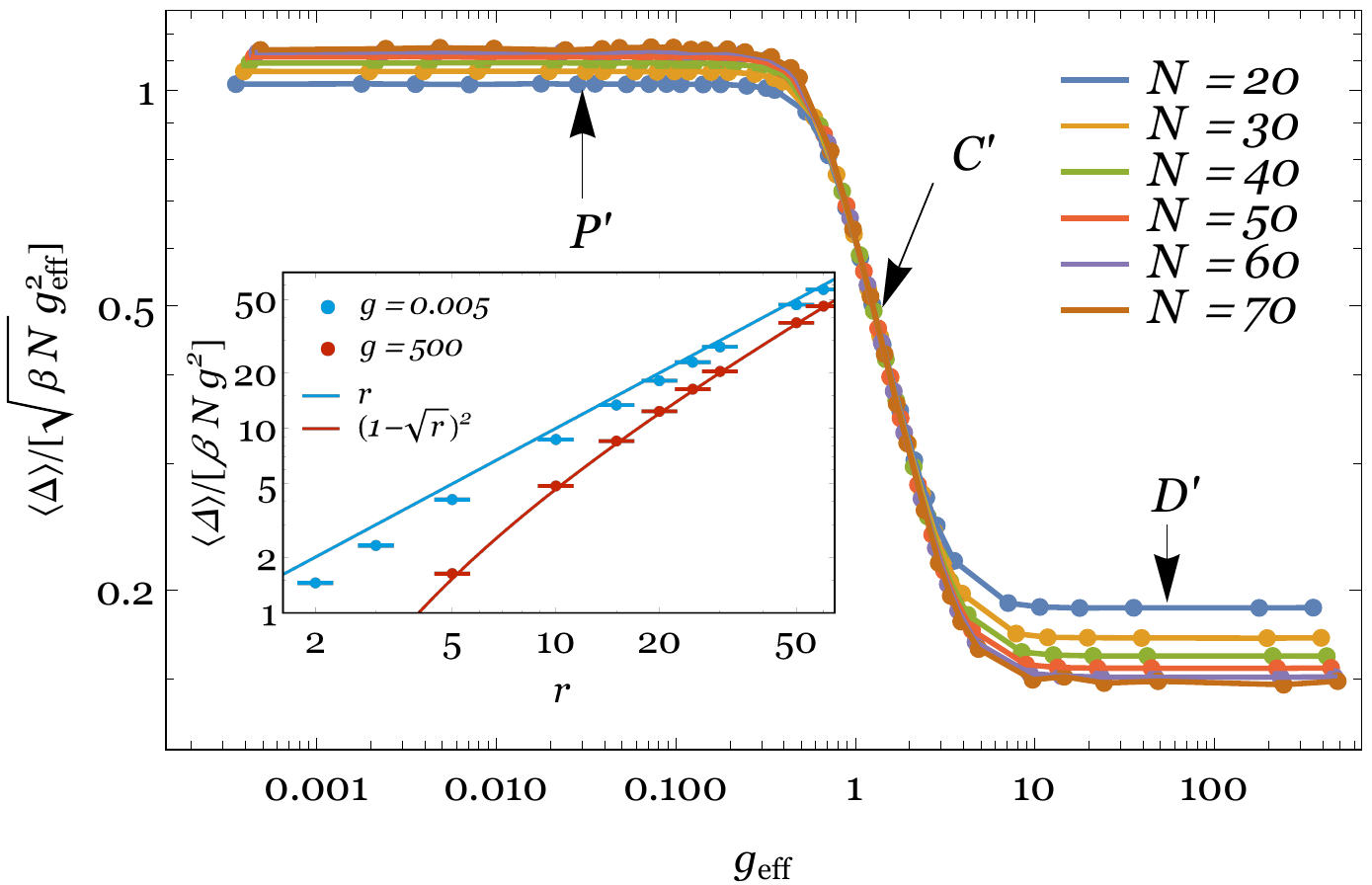}
\caption{\label{fig:gap}Average spectral gap as a function of $g_{\mathrm{eff}}$ plotted for different values of $N$ for $\beta=2$ and $r=2$. Inset: evolution of the spectral gap with the number of jump operators $r$, for $N=60$, $\beta=2$ and $g=0.005$ (blue) and $g=500$ (red); the full lines correspond to the analytic predictions.}
\end{figure}

We now turn to the study of $\Delta$, which is a particularly important spectral feature since it determines the long-time relaxation asymptotics. Figure~\ref{fig:gap} shows the average spectral gap, $\langle\Delta\rangle$, for $\beta=2$ and $r=2$ as a function of $g_{\mathrm{eff}}$ for different values of $N$. In \ref{app:gap}, we checked that in the thermodynamic limit the distribution of the gap becomes sharply peaked around its mean; hence, the latter accurately describes the long-time dynamics.

Here, we also find three qualitatively different regimes (P$'$, C$'$, D$'$) whose boundaries \emph{do not} coincide with those in Figure~\ref{fig:spectrum}. Here the boundaries $g_{\mathrm{eff}}\simeq g_{\mathrm{P'C'}}$ and $g_{\mathrm{eff}}\simeq g_{\mathrm{C'D'}}$, separating the P$'$ and C$'$, and the C$'$ and D$'$ regimes, respectively, are independent of $N$ for $N\to\infty$. For regime P$'$, the average gap behaves as $\langle\Delta\rangle\propto(\beta N)^{1/2}g_{\mathrm{eff}}^{2}$; for C$'$, the gap varies as $\langle\Delta\rangle\propto(\beta N)^{1/2}g_{\mathrm{eff}}$; for D$'$, we observe again $\langle\Delta\rangle\propto(\beta N)^{1/2}g_{\mathrm{eff}}^{2}$. As before, these results are only possible to establish by performing a large-$N$ extrapolation of the available data due to the presence of large finite-size corrections (see \ref{app:gap}).

The limiting $N\to\infty$ values of the spectral gap for small and large $g_{\mathrm{eff}}$ can be determined either by perturbative arguments or by computing the holomorphic Green's functions, respectively (see \ref{app:gap} for details; see also \cite{sa2019thesis,can2019} for a computation for arbitrary $g_\mathrm{eff}$). 
One finds that $\langle\Delta\rangle=\beta Ng^{2}(1-\sqrt{r})^{2}$ for large $g_{\mathrm{eff}}$, which is compatible with~\cite{can2019}. 
For small $g_\mathrm{eff}$, employing degenerate perturbation theory to generalize the results of~\cite{timm2009} (see \ref{app:gap} for details), we find that for large $r$ the average gap is given by $\langle\Delta\rangle=\beta Ng^{2}r$. 
The predictions describe the gap increasingly well for growing $r$; for $r=2$, although not exact, they give a good estimate, see inset in Figure~\ref{fig:gap}. Notwithstanding that the above two results are derived in the limits $g_{\mathrm{eff}}\to0$ and $g_{\mathrm{eff}}\to\infty$, they provide a remarkable description for the whole P$'$ and D$'$ regimes, respectively. For the special case $r=1$, although three regimes are also present (see~\ref{app:r=1}), the scaling of the gap with $N$ changes in the strongly dissipative regime. Finally, none of the regimes above show the mid-gap states reported in~\cite{can2019a}, where only Hermitian jump operators were considered.

\section{Steady-state properties}
\label{sec:steady_state}

\begin{figure}[tbp]
\centering \includegraphics[width=\textwidth]{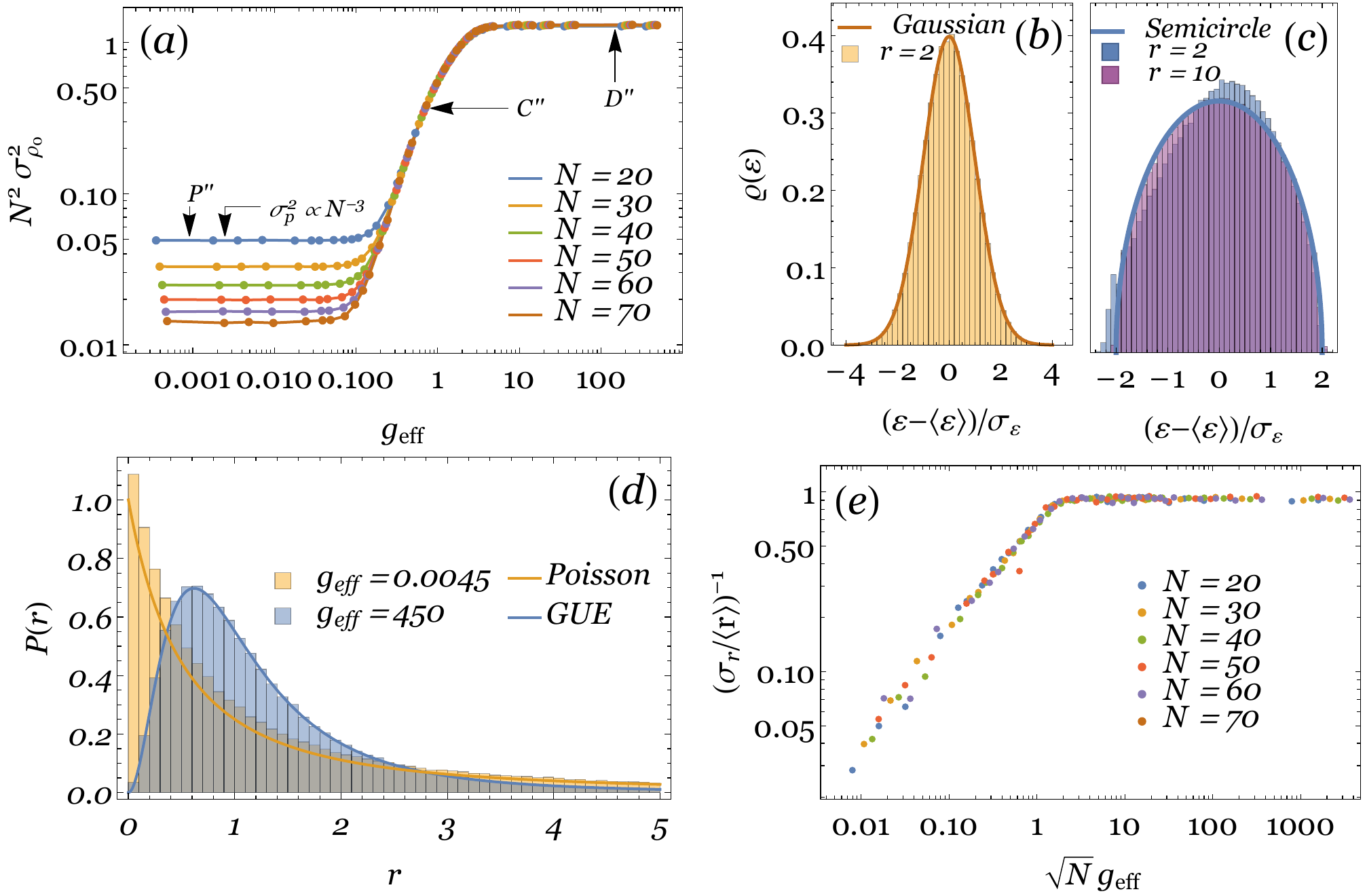}
\caption{$(a)$: Variance of steady-state probabilities as a function of $g_{\mathrm{eff}}$ plotted for different values of $N$ for $\beta=2$ and $r=2$. $(b)-(c)$: spectral density for the effective Hamiltonian $\mathcal{H}$ for $g_\mathrm{eff}=0.0045$ and $g_\mathrm{eff}=450$, respectively. $(d)$: statistics of level spacing ratios for weak and strong dissipation and comparison with approximate analytic predictions for Poisson and RMT statistics (solid lines). $(e)$: ratio of the first two moments of the distribution of $r$, $(\sigma_{r}/\langle r\rangle)^{-1}$, for $\beta=2$ and $r=2$.}
\label{fig:steady_state}
\end{figure}

We next characterize the steady state $\rho_{0}$. First, we consider the variance, $\sigma_{\rho_{0}}^{2}$, of the eigenvalues of $\rho_{0}$. This quantity is related to the difference
between the purity of the steady state, $\mathcal{P}_{0}=\Tr\left(\rho_{0}^{2}\right)$, which quantifies the degree of mixing of $\rho_{0}$, and that of a fully-mixed state $\mathcal{P}_{\mathrm{FM}}=1/N$, $\mathcal{P}_{0}-\mathcal{P}_{\mathrm{FM}}=N\sigma_{\rho_{0}}^{2}$.
Figure~\ref{fig:steady_state}~$(a)$ shows the variance~$\sigma_{\rho_{0}}^{2}$ as a function of $g_{\mathrm{eff}}$ for $\beta=2$ and $r=2$. Here again, three different regimes can be observed whose boundaries \emph{do not} coincide with those given for previous quantities. In regime P$''$, $g_{\mathrm{eff}}\lesssim (\beta N)^{-\frac{1}{2}}$, we observe 
\begin{equation}
    \sigma_{\rho_{0}}^{2}\propto N^{-3}f_{\rho_{0}}^{<}\left(N^{1/2}g_{\mathrm{eff}}\right),
\end{equation}
with $f_{\rho_{0}}^{<}(x)\propto x^{0}$, for $x\!\to\!0$, and $f_{\rho_{0}}^{<}(x)\!\propto\!x^{2}$ for $x\!\to\!\infty$. In regime D$''$, for $N^{0}\lesssim g_{\mathrm{eff}}$, we have 
\begin{equation}
    \sigma_{\rho_{0}}^{2}\propto N^{-2}f_{\rho_{0}}^{>}(g_{\mathrm{eff}}),
\end{equation}
with $f_{\rho_{0}}^{>}(x)\!\propto\!x^{0}$ for $x\!\to\!\infty$ and $f_{\rho_{0}}^{<}(x)\!\propto\!x^{2}$ for $x\!\to\!0$. The crossover regime C$''$ can be accessed by both asymptotic expansions and corresponds to $\sigma_{\rho_{0}}^{2}\propto N^{-2}g_{\mathrm{eff}}^{2}$. The asymptotic matching again agrees with data extrapolation (see \ref{app:steady-state}). These scalings imply that, up to subleading $1/N$ corrections, the steady state is fully mixed in regime P$''$, $\langle\mathcal{P}_{0}\rangle=\mathcal{P}_{\mathrm{FM}}+\mathcal{O}\left(1/N^{2}\right)$, while in regimes C$''$ and D$''$, $\mathcal{P}_{0}$ is only proportional to $\mathcal{P}_{\mathrm{FM}}$. At large dissipation, the steady state can be very well described by a random Wishart matrix, in agreement with general results of the entanglement spectrum of random bipartite systems~\cite{zyczkowski2001,sommers2004}. Purity can then be computed straightforwardly. This will be discussed for a more tractable model in a future publication~\cite{sa2020prep}. The case with $r=1$ is again qualitatively different from $r>1$ at strong dissipation, the purity $\langle\mathcal{P}_0\rangle$ being of order $N^0$ (see \ref{app:r=1}), signaling a steady state closer to a pure state than to a fully-mixed one. 

Next, we investigate the steady state's spectrum. For that, it is useful to introduce the effective Hamiltonian $\mathcal{H}=-\log\rho_{0}$. We denote the eigenvalues of $\mathcal{H}$ by $\varepsilon_i$ and present their spectral density $\varrho(\varepsilon)$ in Figures~\ref{fig:steady_state}~$(b)$ and $(c)$, corresponding to a point in regime P$''$ and D$''$, respectively. In the weak coupling regime, P$''$, $\varrho(\varepsilon)$ is well described by a Gaussian, while at strong coupling, D$''$, it acquires a non-Gaussian shape. For large $r$, $\varrho(\varepsilon)$ is increasingly well described by a Wigner semicircle distribution. A remarkable agreement can already be seen for $r=10$ in the example of Figure~\ref{fig:steady_state}~$(c)$. For small $r$ (see $r=2$ in Figure~\ref{fig:steady_state}~$(c)$) there is a systematic skewing of the spectrum to the right. Figure~\ref{fig:steady_state}~$(d)$ presents the probability distribution, $P(r)$, of adjacent spacing ratios, $r_{i}=s_{i}/s_{i-1}$, with $s_{i}=\varepsilon_{i+1}-\varepsilon_{i}$, which automatically unfolds the spectrum of $\mathcal{H}$~\cite{atas2013}. The analytic predictions for the GUE and for the Poisson distribution~\cite{atas2013} (full lines) are given for comparison. The agreement of the numerical data of the points in the P$''$ and D$''$ regimes, respectively, with the Poisson and GUE predictions is remarkable. Within regime C$''$, we observe a crossover between these two regimes.

To illustrate the crossover in the spectral properties of $\mathcal{H}$ with $g_{\mathrm{eff}}$, we provide in Figure~\ref{fig:steady_state}~$(e)$ the ratio of the first two moments of the distribution of $r$, $(\sigma_{r}/\langle r\rangle)^{-1}$ which can distinguish between Poisson and GUE statistics. Since in the Poissonian case $P(r)=1/(1+r)^{2}$, the $n$-th moment of the distribution diverges faster than the $(n-1)$-th and thus $\sigma_{r}/\langle r\rangle\to\infty$. On the other hand, for the GUE this ratio is given by a finite number of order unity, $\sigma_{r}^2/\langle r\rangle^2=256\pi^{2}/(27\sqrt{3}-4\pi)^{2}-1\simeq1.160$. Figure~\ref{fig:steady_state}~$(e)$ shows that the Poissonian statistics are only attained in the dissipationless limit $N^{1/2}g_{\mathrm{eff}}\to0$. On the other hand, the GUE values are attained for $g_{\mathrm{eff}}N^{1/2}\simeq1$. Thus, in the thermodynamic limit, the effective Hamiltonian $\mathcal{H}$ is quantum chaotic for all finite values of $g_{\mathrm{eff}}$.

\section{Conclusions}
\label{sec:conclusion}

In summary, by analysing an ensemble of stochastic Liouvillians of Lindblad form, where unitary dynamics coexists with $r$ independent dissipation channels, we find that the dispersion of the decay rates, the spectral gap, and the steady-state properties are divided into weak-dissipation, crossover and strong-dissipation regimes, whose boundaries do not necessarily coincide for the different observables. 
For a given observable, each regime and its boundaries are characterised by a set of scaling exponents ruling the dependence on the effective dissipation strength, $g_{\mathrm{eff}}$, and system size $N$. 
We determine these exponents and find them to be independent of the universality index $\beta$ of random matrices. It is important to note that the different scaling regimes refer to the value of the effective coupling constant $g_\mathrm{eff}$. Since $g_{\mathrm{eff}}=(2 r \beta N)^{1 / 4} g$, our large-$g_{\mathrm{eff}}$ regime is, in fact, what should be typically observed in the thermodynamic limit, for any constant finite $g$~\cite{sa2019complex}.

As dissipation increases, the support of the spectrum passes from an ellipse to the lemon-like shape reported in~\cite{denisov2018} for the case of $r=N^{2}-1$. 
For fixed $g_{\mathrm{eff}}$ and $r>1$, the spectral gap increases with $N$ and its distribution becomes peaked in the $N\to\infty$ limit. 
The case of a single dissipation channel, $r=1$, is qualitatively different. For $r=1$, $\Delta$ vanishes with increasing $N$ for infinite dissipation. Finally, with increasing system size and $r>1$, the steady-state purity approaches that of the maximally mixed state, $\mathcal{P}_{\mathrm{FM}}=1/N$, in the week dissipative regime, while for strong dissipation it attains a value larger than, though proportional to, $\mathcal{P}_{\mathrm{FM}}$. Interestingly, the steady-state spectral statistics exhibit a crossover from Poissonian to GUE as a function of $g_{\mathrm{eff}}N^{1/2}$. 

The generic features identified can be contrasted to those of integrable (regular) systems. The lemon-like shape at large dissipation is generic (\cite{denisov2018} discusses the independence from specific sampling schemes), with a universal crossover to an ellipse-shaped spectral support. For integrable systems, the spectral density is highly model-dependent; furthermore, there are, in general, lines of eigenvalues of high degeneracy emanating from the bulk spectrum. The correlations inside the spectrum are also markedly different, according to a generalisation of the quantum chaos conjecture~\cite{sa2019complex}. The steady-state of integrable systems is, furthermore, expected to be Poisson-like for an extensive range of parameters, while in random states it is highly suppressed in the thermodynamic limit. We leave a more detailed comparison for future work.

A natural question our results raise is---which regime characterises a particular physical system? As in the case of chaotic Hamiltonian dynamics, this has to be determined on a case-by-case basis. One must also check whether any non-Markovian effects arise. We delegate these questions to future works.

Finally, an important direction of further research is the extension of the results on random Liouvillians to other symmetry classes. We note that the Altland-Zirnbauer tenfold classification of (closed) Hamiltonian dynamics~\cite{altland1997} has to be extended for non-Hermitian Hamiltonians, leading to the 38-fold classification of Bernard and LeClaire~\cite{bernard2002,kawabata2019}. The symmetries of Liouvillian dynamics restrict the allowed symmetric classes for quadratic Liouvillians back to ten, agreeing with the Altland-Zirnbauer classes in the dissipationless limit~\cite{lieu2020}. A Liouvillian in the new non-standard symmetry class AI$^\dagger$ was recently reported in~\cite{hamazaki2019}. This opens the door for novel effects in open quantum systems to arise in these symmetry classes, with potentially high impact in condensed matter and optical setups.

\appendix

\section*{Appendices}

In the Appendices we provide a summary of the various scaling exponents (for scaling with $N$ and $g_\mathrm{eff}$) for $r>1$ (given in the main text for $r=2$) and for $r=1$ (\ref{app:exponents}). We also give further details for the evolution of global spectral properties (\ref{app:spectrum}), the spectral gap (\ref{app:gap}) and steady-state properties (\ref{app:steady-state}) for $r>1$ and compare with the results for $r=1$ (\ref{app:r=1}).

\section{Summary of exponents}\label{app:exponents}

The compatibility-of-scaling-function arguments given in the main text for $X$ and $\sigma_{\rho_0}^2$ can be given in general, for any quantity which has the same qualitative behaviour as the ones described in this work. This procedure also allows us to systematize the exponents found in the main text, by defining three types of exponents: $\nu$, $\kappa$, and $\lambda$, related to the finite-size scaling of the spectral and steady-state quantities (say, $X$), the finite-size scaling of the boundaries of the multiple regimes (say, P, C, D) and the $g_\mathrm{eff}$-scaling of the quantities in the crossover regime (say, of $X$ in C), respectively.

We define three regimes with $N$-dependent boundaries in which some quantity $Q$ has qualitatively distinct behaviours: P$_Q$, for $g_\mathrm{eff}\lesssim(\beta N)^{\kappa_Q^<}$; C$_Q$ for $(\beta N)^{\kappa_Q^<}\lesssim g_\mathrm{eff}\lesssim(\beta N)^{\kappa_Q^>}$; and D$_Q$ for $(\beta N)^{\kappa_Q^>}\lesssim g_\mathrm{eff}$. In P$_Q$, $Q$ behaves as 
\begin{equation}
    Q\propto g_\mathrm{eff}^2(\beta N)^{\nu_Q^P}f_Q^<[(\beta N)^{-\kappa_Q^<}g_\mathrm{eff}]\,,
\end{equation} while in D$_Q$,
\begin{equation}
    Q\propto g_\mathrm{eff}^2(\beta N)^{\nu_Q^D}f_Q^>[(\beta N)^{-\kappa_Q^>}g_\mathrm{eff}]\,.
\end{equation}
Note that some extra factors of $g_\mathrm{eff}$ may exist (such as the extra $g_\mathrm{eff}^2$ for $X$) but they do not modify the argument, as long as they are the same in regimes P$_Q$ and D$_Q$. If they are different, a straightforward modification of (\ref{eq:exponents_constraint}) below is required, but this issue did not arise for the quantities studied in this work. We further assume that the scaling functions $f_Q$ have asymptotic power-law behaviours, that is, 
\begin{equation}
    f_Q^<(x)\!\propto\! x^0\quad\mathrm{ if }\quad x\!\to\!0,
\end{equation}
\begin{equation}
    f_Q^<(x)\!\propto\! x^{\lambda_Q^<}\quad \mathrm{if} \quad x\!\to\!\infty,
\end{equation}
\begin{equation}
    f_Q^>(x)\!\propto\! x^0\quad \mathrm{if}\quad x\!\to\!\infty,
\end{equation}
\begin{equation}
    f_Q^>(x)\!\propto\! x^{\lambda_Q^>}\quad \mathrm{if}\quad x\!\to\!0.
\end{equation}
The two limiting behaviours of $Q$ should match in the intermediate regime C$_Q$, that is,
\begin{equation}
\fl
    \lim_{g_\mathrm{eff}(\beta N)^{-\kappa_Q^<}\to0}(\beta N)^{\nu_Q^P}f_Q^<[(\beta N)^{-\kappa_Q^<}g_\mathrm{eff}]
    \simeq\lim_{g_\mathrm{eff}(\beta N)^{-\kappa_Q^>}\to\infty}(\beta N)^{\nu_Q^D}f_Q^>[(\beta N)^{-\kappa_Q^>}g_\mathrm{eff}]\,,
\end{equation}
whence the equality
\begin{equation}
    g_\mathrm{eff}^{2+\lambda_Q^<}\left(\beta N\right)^{\nu_Q^P-\kappa_Q^<\lambda_Q^<}=g_\mathrm{eff}^{2+\lambda_Q^>}\left(\beta N\right)^{\nu_Q^D-\kappa_Q^>\lambda_Q^>}
\end{equation}
follows. This equality implies that $\lambda_Q^<=\lambda_Q^>\equiv\lambda_Q$ and establishes a relation between $\nu$, $\kappa$ and $\lambda$ exponents, which are thus not all independent but constrained by
\begin{equation}\label{eq:exponents_constraint}
    \lambda_Q\left(\kappa_Q^>-\kappa_Q^<\right)=\nu_Q^D-\nu_Q^P\,.
\end{equation}

Table~\ref{tab:exponents} shows the values of the various exponents for $r=1$ and $r>1$, as defined above. It can be checked that all satisfy (\ref{eq:exponents_constraint}). Note that the exponent $\nu_Q^C$ (giving the collapse of curves of different $\beta N$ in regime $C_Q$) is not defined in the above argument and thus does not enter the constraint.

\begin{table}[t]
\caption{\label{tab:exponents}Scaling exponents for $r=1$ and $r>1$, as defined in the main text. $\nu$ exponents give the finite-size scaling of the spectral and steady-state quantities within one of the regimes P, C, D and their primed counterparts; $\lambda$ exponents give the extra $g_\mathrm{eff}$-scaling besides $g_\mathrm {eff}^2$ in regime C and its primed counterparts; $\kappa$ exponents give the finite-size scaling of the boundaries of the regimes.}
\vspace{+0.5em}
\begin{tabular}{r|rrrrrr|rrrrrr}
$r$ &
$\nu_P$ &
$\nu_C$ &
$\nu_{D}$ &
$\lambda_X$ &
$\kappa_X^<$ &
$\kappa_X^>$ &
$\nu_{P'}$ &
$\nu_{C'}$ &
$\nu_{D'}$ &
$\lambda_{\Delta}$ &
$\kappa_{\Delta}^<$ &
$\kappa_{\Delta}^>$ \\
\hline\hline
$=1$  & $0$ & $1/2$ & $1/2$ & $2$ & $-1/4$ & $0$ & $1/2$ & $1/2$ & $-3/2$ & $-8/3$ & 0 & $3/4$\\
$>1$  & $0$ & $1/2$ & $1/2$ & $2$ & $-1/4$ & $0$ & $1/2$ & $1/2$ & $1/2$  & ---    & $0$ & $0$\\
\end{tabular}
\vspace{+1.5em}
\begin{tabular}{r|rrrrrr}
$r$ &
$\nu_{P''}$ &
$\nu_{C''}$ &
$\nu_{D''}$ &
$\lambda_{\rho_0}$ &
$\kappa_{\rho_0}^<$ &
$\kappa_{\rho_0}^>$\\
\hline\hline
$=1$  & $-3$ & $-2$ & $-1$ & $8/5$ & $-1/2$ & $3/4$\\
$>1$  & $-3$ & $-2$ & $-2$ & $2$   & $-1/2$ & $0$   \\
\end{tabular}
\end{table}

\section{Global Spectrum}\label{app:spectrum}

\begin{figure}[tbp]
\centering
    \includegraphics[width=0.65\textwidth]{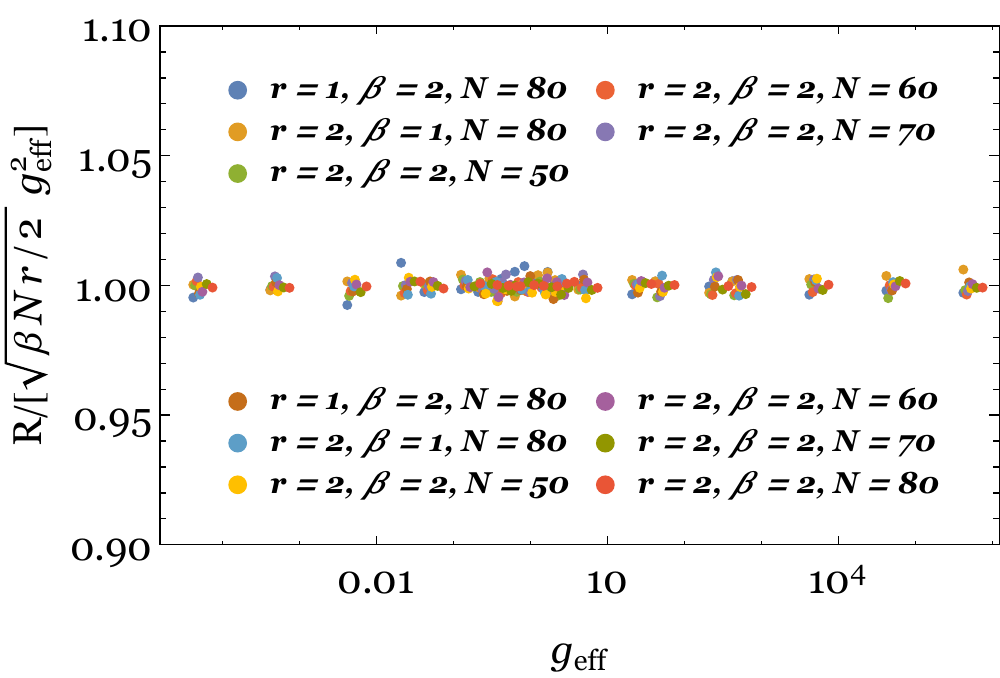}
    \caption{Center of mass $R$ of the spectrum for different combinations of $N$, $\beta$, $r$ and $g$.}
    \label{fig:SM_R}
\end{figure}

The scaling of spectral quantities can be motivated as follows. The spectral density of $H$ is given by the Wigner semicircle law, with zero mean and standard deviation $\sqrt{\beta N/2}$. The dissipation matrix $d$ is a sum of $r$ Wishart matrices, hence its spectral density follows a Marchenko-Pastur law, with mean $\beta N r g^2$ and standard deviation $\sqrt{r}\beta N g^2$. Since the eigenvalues of $H$ and $d$ are real, when considered separately, i.e.\ at the non-dissipative ($g=0$) or at the fully dissipative ($g=\infty$) limits, respectively, they only contribute to the imaginary or to the real parts of $\Lambda$. Although at finite $g$ these considerations are no longer exact, we expect them to yield the leading scaling behaviour. Thus, only the dissipative term contributes to the mean of $\Lambda$ and we have $R \propto \beta N r g^2$, see Figure~\ref{fig:SM_R}. At large $g$, the main contribution to $X$ comes from the dissipative term and $X\propto \sqrt{r}\beta N g^2$ (see Figure~\ref{fig:spectrum}~$(d)$ in the main text). For weak dissipation the $Y$ scaling is dominated by the Hamiltonian term, hence $Y\propto\sqrt{\beta N}$ (see Figure~\ref{fig:spectrum}~$(e)$ in the main text). The passage from the weak dissipation scalings to the strong dissipation scalings should occur when the two terms in the Liouvillian are of the same order, $\sqrt{\beta N/2}\simeq \beta N \sqrt{r} g^2$, i.e.\ $g_\mathrm{eff}\simeq 1$.

\begin{figure}[tbp]
\centering
    \includegraphics[width=0.99\textwidth]{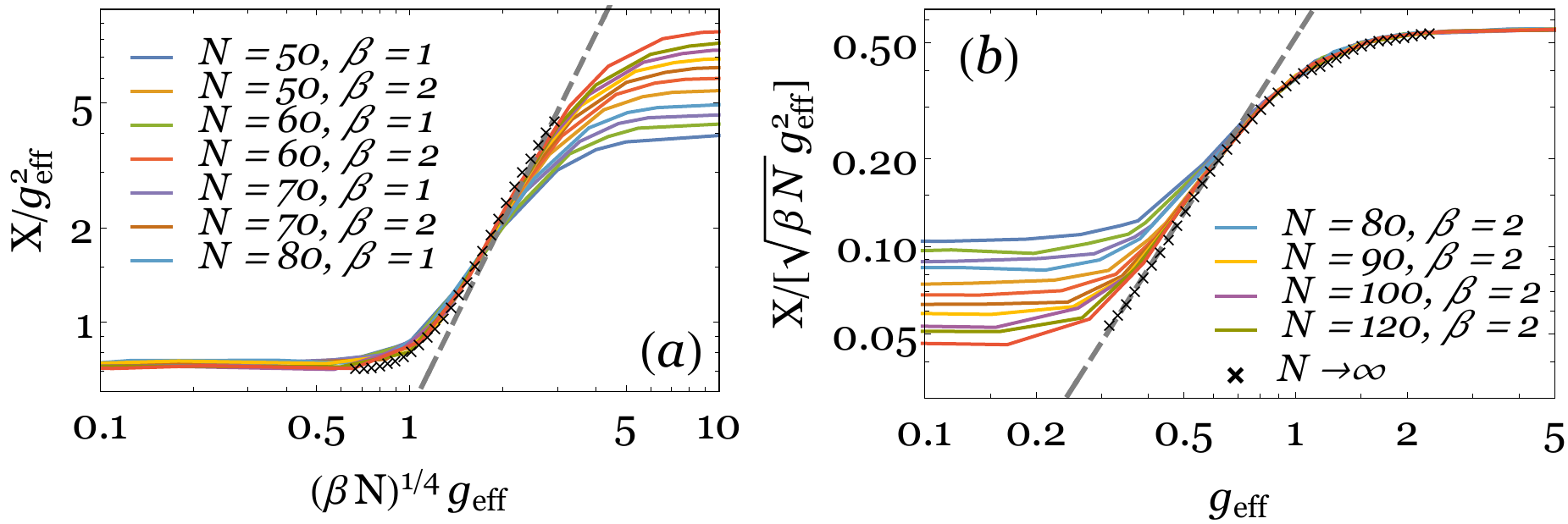}
    \caption{$X$ as a function of $g_\mathrm{eff}$ for $r=2$ and various $N$ and $\beta$. The black crosses give the extrapolation of the data for $N\to\infty$ and the dashed gray line is $\propto g_\mathrm{eff}^{\lambda_X}$. Collapse to the universal curve for small $g_\mathrm{eff}$ in $(a)$ and for large $g_\mathrm{eff}$ in $(b)$.}
    \label{fig:SM_X_ScalingFunc} 
\end{figure}

By fitting the global spectral data to power-laws, we find very good agreement with the exponents in Table~\ref{tab:exponents}. $\nu_C=\nu_D$ is completely independent of $r$. $\nu_P$ seems to decrease slightly with $r$, although the dependence is weak, and is most likely due to the reduced sizes possible to attain numerically. Furthermore, the exponents $\lambda_X=2$ and $\kappa_X^<=-1/4$ are compatible with data extrapolation for $N\to\infty$, see Figure~\ref{fig:SM_X_ScalingFunc}. For both large $g_\mathrm{eff}$ (Figure~\ref{fig:SM_X_ScalingFunc}~$(a)$) and small $g_\mathrm{eff}$ (Figure~\ref{fig:SM_X_ScalingFunc}~$(b)$), the extrapolated data to $N\to\infty$ (black crosses), taken as the $y$-intercept of a linear fit in $(\beta N)^{-1}$, agrees very well with the power-law $g_\mathrm{eff}^{\lambda_X}$ (gray line) in regime C. Upon entering regimes P and D, the extrapolation points naturally start deviating from the power-law curve.

\section{Spectral Gap}\label{app:gap}

\begin{figure}[tbp]
\centering 
\includegraphics[width=0.99\textwidth]{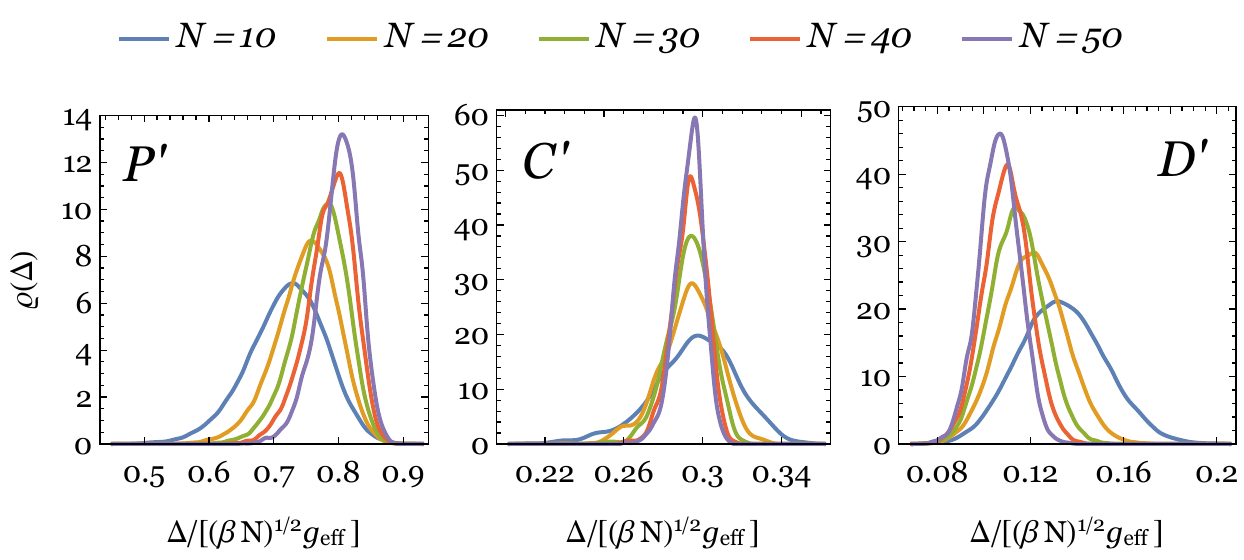}
\caption{\label{fig:SM_gap_distribution_function}Scaled gap distribution function for each of the regimes P$'$, C$'$, D$'$.}
\end{figure}

We now consider the spectral gap for $r>1$. For all regimes, the variance of the gap decreases with $N$ and the value of $\Delta$ becomes sharply defined around its mean, see the gap distribution functions depicted in Figure~\ref{fig:SM_gap_distribution_function}. Hence, the mean gap accurately describes the long-time dynamics. 

The mean of the different distributions in Figure~\ref{fig:SM_gap_distribution_function} is not constant due to the finite-size effects, which we address next. Fitting directly the numerical average spectral gap for several $N$ and multiple $r>1$ seems to indicate that there is a dependence of the scaling exponents $\nu_{P'}$, $\nu_{C'}$, $\nu_{D'}$ with $r$. In particular, $\nu_{P'}=\nu_{C'}=\nu_{D'}$ only for large $r$ but $\nu_{P'}>\nu_{C'}>\nu_{D'}$ for small $r>1$. However, this is incompatible with both the picture drawn in \ref{app:exponents} and with the finiteness of the gap in the thermodynamic limit for $r>1$, discussed below. Since we can only access relatively small $N$, this apparent $r$-dependence is likely a finite-size effect suppressed by $r$. Indeed, values of the exponents obtained by data extrapolation to large $N$ are compatible with $\nu_{P'}=\nu_{C'}=\nu_{D'}=1/2$, independently of $r$. 

We next provide the analysis yielding the results presented in the main text on the limiting behaviour of the spectral gap at small and large $g_\mathrm{eff}$. For strong dissipation, we have $\mathcal{L}\simeq\sum_\ell\mathcal{D}_{W_\ell}$. The endpoints of the spectrum of $\mathcal{L}$ along the real direction in the complex plane (that is, the spectral gap and left-most point of the lemon-shaped support, which is the gap plus some order-unity multiple of $X$) can be accessed directly by the use of the holomorphic Green's function of $\mathcal{L}$~\cite{feinberg1997,jurkiewicz2008}, $G=\Tr\langle(z-\mathcal{L})^{-1}\rangle$.
This was exploited in~\cite{can2019} to compute the spectral form factor.
Unfortunately, the information along the imaginary axis (for instance $Y$) and the spectral density inside the support cannot be accessed by this method.

When computing the holomorphic Green's function, it can be shown~\cite{can2019,sa2019thesis} that the terms $W_\ell\otimes W^*_\ell$ do not contribute in the large-$N$ limit. For a single jump operator, the superoperator under consideration is then of the form $\mathcal{L} \simeq -\frac{1}{2}(\Gamma\otimes\mathbbm{1}+\mathbbm{1}\otimes\Gamma$), where $\Gamma=W^\dagger W$ is a square Wishart matrix. Since the two sectors of the Liouvillian tensor product representation do not see each other, we can do independent expansions for each Wishart matrix. Then,  $n^\mathrm{th}$-order terms of the expansion of the Liouvillian Green's function are given by all possible combinations of $n_1^\mathrm{th}$-order terms of one of the Wishart matrices with all $n_2^\mathrm{th}$-order terms of the other Wishart matrix, restricted by $n_1+n_2=n$ and there are ${n}\choose{n_1}$ such combinations for each fixed $n$. By summing all terms, the holomorphic Green's function of $\mathcal{L}$ is then
\begin{equation}\label{eq:Liouvillian_G_series}
    G_\mathcal{L}(z)=\sum_{n=0}^\infty\frac{1}{z^{n+1}}\left(-\frac{1}{2}\right)^2\sum_{k=0}^n{{n}\choose{k}}C_k\,C_{n-k}\,,
\end{equation}
where $C_k$ counts the number of Wishart planar diagrams at $k^\mathrm{th}$ order and is given by the Catalan numbers, whose real-integral representation is $C_n=(1/2\pi)\int_0^4\d x\,x^{n-1}\sqrt{4x-x^2}$, i.e.\ the moments of the Marchenko-Pastur distribution, $\varrho_{\mathrm{MP}}(x)=\sqrt{x(4-x)}/(2\pi x)$, $0<x<4$. Inserting the integral representation of the Catalan numbers into (\ref{eq:Liouvillian_G_series}), performing the sum and comparing with the relation between the holomorphic Green's function and the spectral density~$\varrho_\mathcal{L}$, $G_\mathcal{L}(z)=\int\d\nu\,\varrho_\mathcal{L}(\nu)/(z-\nu)$, we arrive at
\begin{equation}\label{eq:Liouvillian_rho_integral}
    \varrho_\mathcal{L}(x)=2\int_{\nu_m}^{\nu_M}\d\nu\,\varrho_{\mathrm{MP}}(\nu)\varrho_{\mathrm{MP}}(-\nu-2x)\,,
\end{equation}
with $\nu_m=\max\{0,-2x-4\}$ and $\nu_M=\min\{4,-2x\}$, whence it follows that $-4<x<0$. Hence, the computation of the holomorphic Green's function has lead us to a convolution of two Marchenko-Pastur laws, which is the spectral density of half the sum of two uncorrelated Wishart matrices. Although the spectral density $\varrho_\mathcal{L}$ does not correctly describe the spectrum of the Liouvillian, it does predict its endpoints and the interval $[-4,0]$ in which $\re(\Lambda)$ is supported. Note, in particular, that the spectrum is gapless for a single decay channel and infinite dissipation.

For more than one jump operator, $\Gamma$ is a sum of $r$ identical (i.e.\ independent but sampled from the same distribution) Wishart matrices. In~\cite{can2019} it was shown, using free-addition of random matrices, that the spectral density of $\Gamma$ is again a Marchenko-Pastur distribution, but with endpoints dependent on $r$,
\begin{equation}
    \varrho_\Gamma(x)=\frac{1}{2\pi x}\sqrt{(\xi_+-x)(x-\xi_-)}\,,\quad \xi_-<x<\xi_+\,,
\end{equation}
where $\xi_{\pm}=(1\pm\sqrt{r})^2$. But we have just seen that the computation of the spectral density from the holomorphic Green's function is equivalent to computing it for the half-sum of two uncorrelated matrices $\Gamma$, whence it immediately follows that $\mathcal{L}$ is supported in $[-\xi_+,-\xi_-]$. We thus see, in agreement with~\cite{can2019}, that \begin{equation}\label{eq:gap_strong_dissipation}
    \langle\Delta\rangle=\beta N g^2\xi_-=\beta N r g^2=\sqrt{\beta N}g_\mathrm{eff}^2\frac{(1-\sqrt{r})^2}{\sqrt{2r}}\,,
\end{equation}
while 
\begin{equation}
    X\propto\beta N g^2(\xi_+-\xi_-)=4\beta N \sqrt{r} g^2\propto\sqrt{\beta N}g_\mathrm{eff}^2\,,
\end{equation}
with the proportionality constant of order unity, and where we have reintroduced the variances $\beta N g^2$. The latter result was already verified in the main text and motivated in an alternative way in the previous section. 

\begin{figure}[tbp]
\centering
    \includegraphics[width=0.65\textwidth]{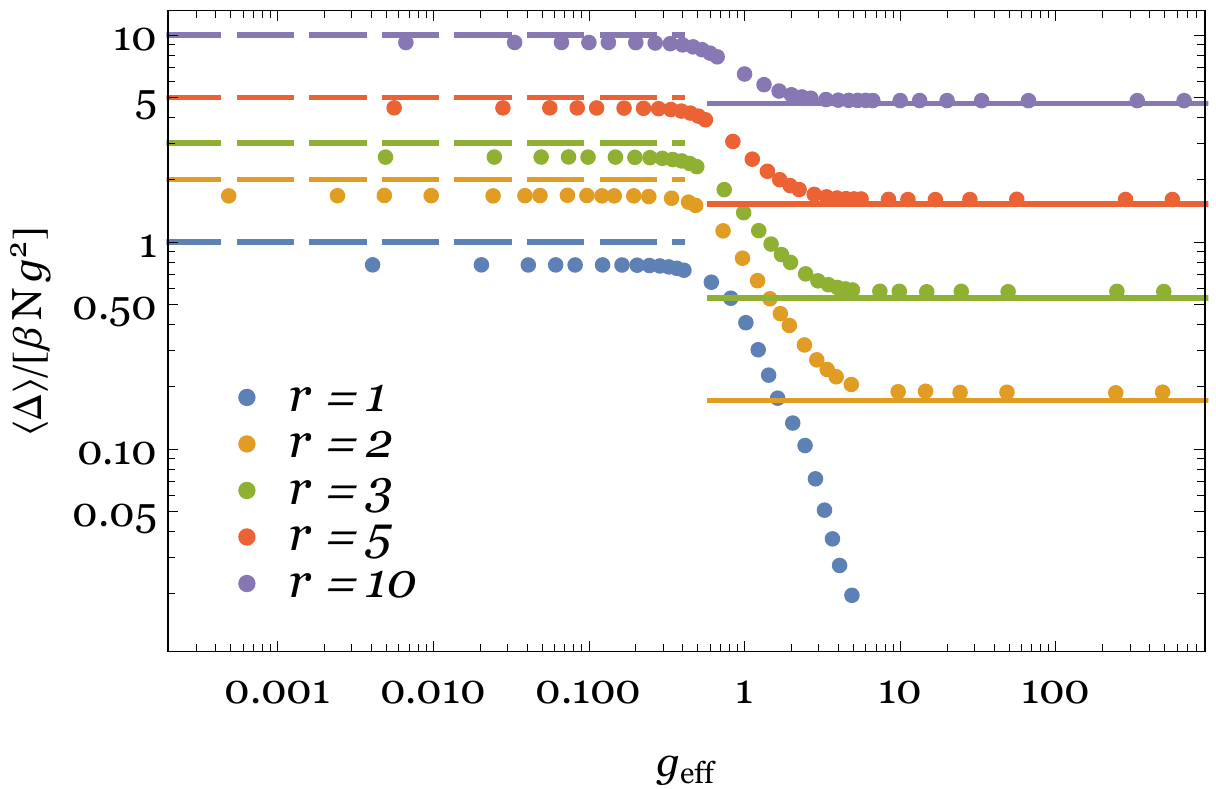}
    \caption{Evolution of the extrapolated spectral gap at $N\to\infty$ with $g_\mathrm{eff}$ for $r=2$, $3$, $5$ and $10$. The the dashed lines are the perturbative results $\langle\Delta\rangle=\beta N g^2 r$ for small $g_\mathrm{eff}$, the full lines the analytic predictions $\langle\Delta\rangle=\beta N g^2(1-\sqrt{r})^2$ for large $g_\mathrm{eff}$.}
    \label{fig:SM_gap_rank}
\end{figure}

The prediction for the gap at large $g_\mathrm{eff}$ for different $r$ is depicted in Figure~\ref{fig:SM_gap_rank}, together with extrapolated data to $N\to\infty$ for several values of $r$, and in Figure~\ref{fig:gap} (main text) for $r$ up to $60$. We see that even at small $r$, the theoretical prediction correctly describes the spectral gap for a wide range of $g_\mathrm{eff}$ (that is, in the whole D$'$ regime). Note also that $\langle\Delta\rangle\to0$ at large $g_\mathrm{eff}$ for $r=1$ in the thermodynamic limit, consistent with our results in \ref{app:r=1}.

For small $g_\mathrm{eff}$, we can proceed perturbatively. If $H$ is diagonal in the basis $\ket{k}$, $H=\sum_k\varepsilon_k\ket{k}\bra{k}$, then the eigenstates of $\mathcal{L}$ at $g=0$ are $\rho_{kk^\prime}=\ket{k}\bra{k^\prime}$ with eigenvalue $\Lambda_{kk^\prime}=-i(\varepsilon_k-\varepsilon_{k^\prime})$ ($k,k^\prime=1,\dots,N$). We thus have a $N$-fold degeneracy of stationary eigenstates, while the other eigenvalues form $N(N-1)/2$ complex conjugate pairs. Any amount of dissipation lifts the degeneracy of the zero-eigenvalue subspace but for small enough dissipation the two sectors will not mix and we expect that the long-time dynamics (the gap) and the (now unique) steady state are determined from the states in the (previously-) degenerate subspace. We lift the degeneracy by diagonalising the perturbation $\sum_\ell\mathcal{D}_{W_\ell}$ in that subspace. We obtain the eigenvalues of $\mathcal{L}$ to first order,
\begin{equation}\label{eq:A_from_LD}
\fl
    A_{nm}\equiv\,\bbra{nn}\sum_\ell\mathcal{D}_{W_\ell}\kket{mm}
    =\left\{  \begin{array}{r@{\quad}cr} 
        \sum_{\ell}W^{(\ell)}_{nm}{W^{(\ell)}_{nm}}^*, &\mathrm{if}&m\neq n\,, \\
        -\sum_{\ell}\sum_{k\neq m}W^{(\ell)}_{km}{W^{(\ell)}_{km}}^*,&\mathrm{if}&m=n\,,
    \end{array}\right.
\end{equation}
with the superoperator notation $\kket{mn}=\ket{m}\otimes\bra{n}^{\sf T}$.

From (\ref{eq:A_from_LD}) it follows that
\begin{equation}
    A_{nm}\in\mathbb{R}\quad (\mathrm{for\ all\ } n)\,,
\end{equation}
\begin{equation}
    A_{nm}>0\quad (n\neq m)\,,
\end{equation}
\begin{equation}
    \sum_{n=1}^N A_{nm}=0\quad (\mathrm{for\ all\ } n)\,,\label{eq:restriction_A}
\end{equation}
which are the conditions for $A$ to be the generator of a classical stochastic equation~\cite{timm2009,denisov2018}, $\partial_t \mathbf{P}=A\mathbf{P}$, where $\mathbf{P}$ is a probability vector. The diagonal entries $A_{mm}=-\sum_{k\neq m}A_{km}$ are determined by the off-diagonal entries, hence we must only determine the distribution of the latter. For $n\neq m$, the distribution of the entry $A_{nm}$ depends on the nature of the jump operator as follows. $W^{(\ell)}_{nm}$ is a Gaussian random variable with variance $g^2$ with $\beta$ real degrees of freedom. Then $A_{nm}=\sum_\ell {W^{(\ell)}_{nm}}^*W^{(\ell)}_{nm}=\sum_\ell\abs{W^{(\ell)}_{nm}}^2$ is (in terms of real degrees of freedom) a sum of $r\beta$ squared Gaussian iid variables and hence follows a $\chi^2$-distribution with $k=r\beta$ degrees of freedom. The distribution function of the entries $A_{nm}$ is 
\begin{equation}\label{eq:distribution_A}
    P_k(A_{nm})=\frac{(A_{nm})^{\frac{k}{2}-1}\exp{-\frac{A_{nm}}{2g^2}}}{(2g^2)^\frac{k}{2}\,\Gamma\left(\frac{k}{2}\right)}\,.
\end{equation}

Ensembles of real exponentially-distributed matrices (corresponding to $k=2$) with the diagonal constraint~(\ref{eq:restriction_A}) were studied in~\cite{timm2009}. In particular, it was shown that, in the large-$N$ limit the spectral density depends only on the second moment of the distribution, $\tau^2_k\equiv\langle A_{nm}^2\rangle-\langle A_{nm}\rangle^2=2kg^4$, and can therefore be calculated from any distribution of entries having this variance, particularly for a Gaussian distribution, which was done in~\cite{staring2003} for symmetric $A$. Furthermore, \cite{timm2009} numerically found that the average smallest nonzero eigenvalue of $A$ (the symmetric value of the spectral gap here) is $N\langle R\rangle$ for large $N$, where $\langle R\rangle$ is the mean of the exponential ($k=2$) distribution of $A_{nm}$. Generalising to our $\chi^2$-distribution, with mean $\mu_k=kg^2$, we find that $\langle\Delta\rangle=\beta N r g^2=\sqrt{\beta N r/2}g_\mathrm{eff}^2$. 

The comparison of this result with numerical data for the average spectral gap for various $r$ is given in Figure~\ref{fig:SM_gap_rank} as a function of $g_\mathrm{eff}$ and for extrapolated $N\to\infty$ and in the inset of Figure~\ref{fig:gap} (main text) for $r$ up to $r=60$ for fixed $N=60$ and two fixed $g$ (one for strong dissipation, one for weak). We see that although there is some deviation from the theoretical curves at small $r$, they describe the numerical data increasingly well with growing $r$.

Finally, we note that the exponent $\lambda_\Delta$ is not defined. The arguments of \ref{app:exponents} do not provide the value of $\lambda_\Delta$ since $\nu_P'=\nu_D'$ and $\kappa_\Delta^<=\kappa_\Delta^>$. This stems from the fact that, for the gap, different scaling functions for small and large dissipation, which asymptotically match in an intermediate regime of size growing with $N$, do not exist. Instead, the gap is described by a single scaling function for all $g$.

\section{Steady state}\label{app:steady-state}

\begin{figure}[tbp]
\centering
    \includegraphics[width=0.99\textwidth]{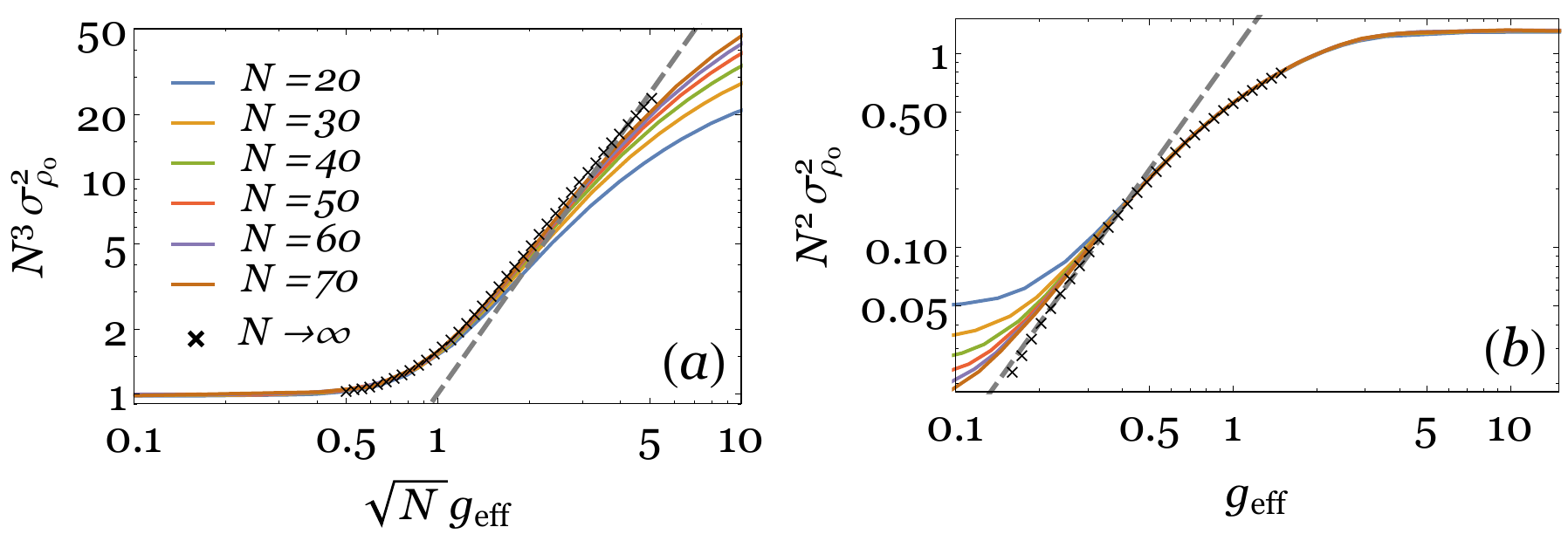}
    \caption{$\sigma_{\rho_0}^2$ as a function of $g_\mathrm{eff}$ for $r=2$ and various $N$. The black crosses give the extrapolation of the data for $N\to\infty$ and the dashed gray line is $\propto g_\mathrm{eff}^{\lambda_{\rho_0}}$. Note data collapse to the universal curve for small $g_\mathrm{eff}$ in $(a)$ and for large $g_\mathrm{eff}$ in $(b)$.}
    \label{fig:SM_purity_ScalingFunc}
\end{figure}

Next, we analyse the steady-state properties for $r>1$. The finite-size scaling exponents $\nu_{P''}=-3$ and $\nu_{C''}=\nu_{D''}=-2$ are in very good agreement with fits of the data to the respective power-laws, for various $r>1$ ($r=2$, $3$, $5$, $10$) and show no dependence on $r$ as expected. The exponents $\lambda_{\rho_0}$ and $\kappa_{\rho_0}^<$ are also compatible with data extrapolation, see Figure~\ref{fig:SM_purity_ScalingFunc}, by proceeding in the same way as for $X$.

\section{Spectral and steady-state properties for \texorpdfstring{$r=1$}{r=1}}\label{app:r=1}

For weak dissipation, the unitary contribution to the Liouvillian dominates and, therefore, the spectral and steady-state properties do not differ qualitatively between $r=1$ and $r>1$. In fact, in regime P$''$, the exponents are always the same in both cases, see Table~\ref{tab:exponents}. However, there are important differences at large $g_\mathrm{eff}$, where dissipation dominates, for both the spectral gap and the steady state. 

\begin{figure}[tbp]
\centering
    \includegraphics[width=0.99\textwidth]{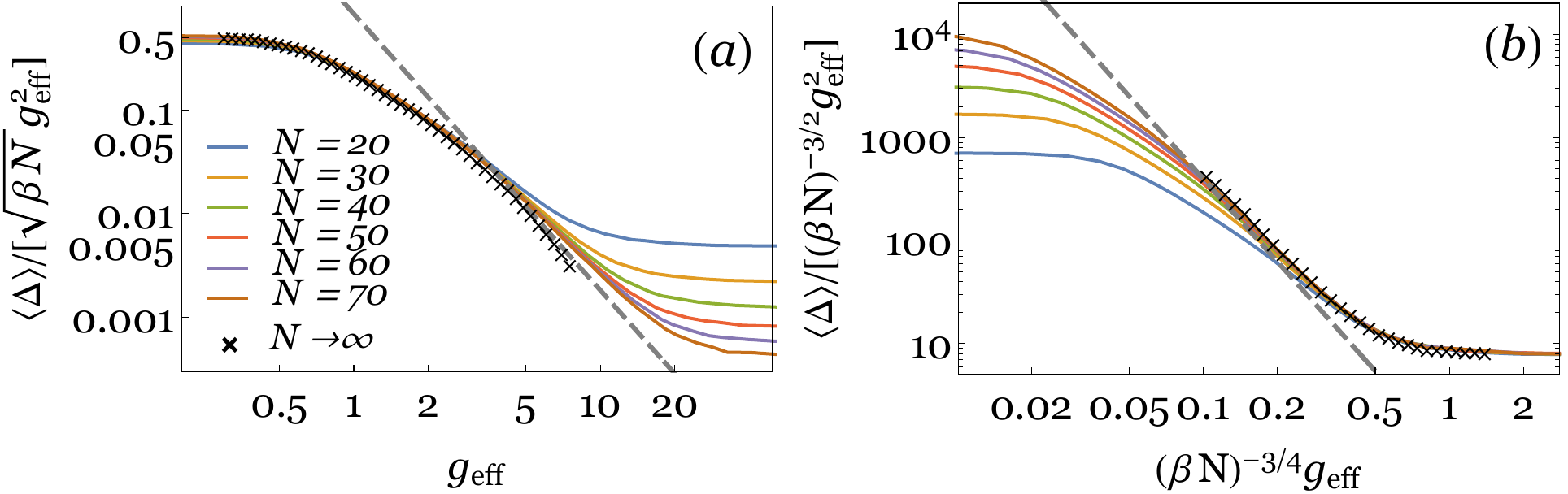}
    \caption{Average spectral gap as a function of $g_\mathrm{eff}$ for $r=1$ and various $N$. The black crosses give the extrapolation of the data for $N\to\infty$ and the dashed gray line is $\propto g_\mathrm{eff}^{\lambda_\Delta}$. Collapse to the universal curve for small $g_\mathrm{eff}$ in $(a)$ and for large $g_\mathrm{eff}$ in $(b)$.}
    \label{fig:SM_r1_gap_ScalingFunc}
\end{figure}

Contrary to $r>1$, for $r=1$ the scaling of the gap at large $g_\mathrm{eff}$ and at small $g_\mathrm{eff}$ is not the same. In particular, since $\nu_{D'}$, $\lambda_\Delta<0$, in the thermodynamic limit the gap starts to close as dissipation increases (strictly closing only for $g_\mathrm{eff}=+\infty$). This can be seen in Figure~\ref{fig:SM_r1_gap_ScalingFunc}, where we plot the evolution of $\langle\Delta\rangle$ as a function of $g_\mathrm{eff}$: choosing the small-$g_\mathrm{eff}$ scaling (which coincides with the scaling for all $g_\mathrm{eff}$ for $r>1$), at large $g_\mathrm{eff}$ the gap follows the dashed line (power-law $g_\mathrm{eff}^{2-\lambda_\Delta}$) and hence closes at $g_\mathrm{eff}=\infty$. Once again, the exponents listed in Table~\ref{tab:exponents} are compatible with data extrapolation, see the black crosses in Figure~\ref{fig:SM_r1_gap_ScalingFunc}. The procedure for data extrapolation is the same as before. 

Furthermore, for the $r=1$ case, contrary to $r>1$, the values of $g_\mathrm{eff}$ that define the boundaries between different regimes are $N$-dependent. Hence, $\lambda_\Delta$ can be defined for $r=1$. The gap has a nonzero exponent $\kappa_\Delta^>$ and the matching procedure of \ref{app:exponents} is applicable. In particular, with $\kappa_\Delta^>=3/4$ (see Figure~\ref{fig:SM_r1_gap_ScalingFunc}~$(b)$), we can use (\ref{eq:exponents_constraint}) to determine $\lambda_\Delta=-8/3$. Hence, we find, at large dissipation, $\langle\Delta\rangle\propto g_\mathrm{eff}^{-2/3}$ in agreement with~\cite{can2019}.

\begin{figure}[t]
\centering
    \includegraphics[width=0.99\textwidth]{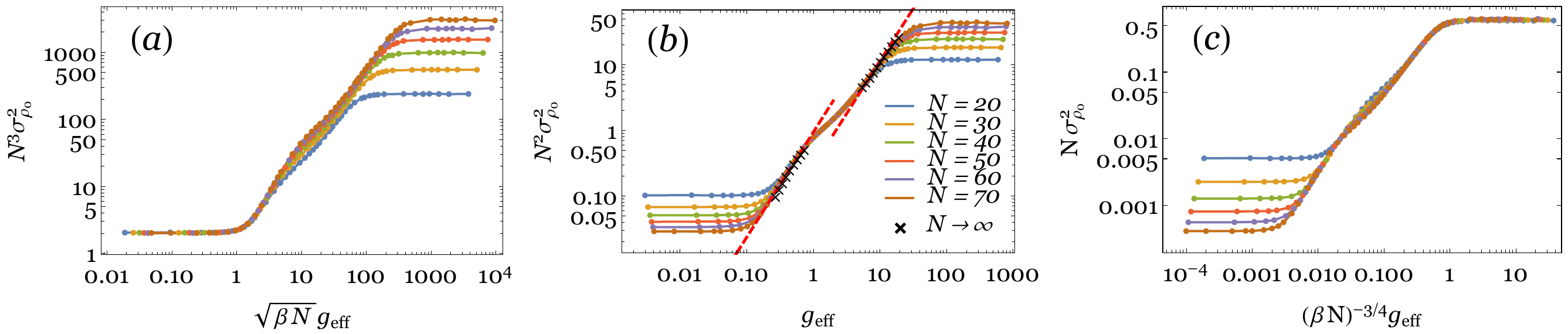}
    \caption{Variance of steady-state probabilities as a function of $g_{\mathrm{eff}}$ plotted for a single decay channel ($r=1$), different values of $N$ and $\beta=2$. The rescalings of $\sigma^2_{\rho_0}$ and of $g_\mathrm{eff}$ are those so as to collapse the different curves to a single curve in the $(a)$ P$''$ regime, $(b)$ C$''$ regime, and $(c)$ D$''$ regime. The dashed lines in $(b)$ are obtained by fitting extrapolated data for $N\to\infty$ (black crosses) to power-law behaviour $g_\mathrm{eff}^{\lambda_{\rho_0}}$ near the boundaries of regime C$''$.}
    \label{fig:SM_r1_purity}
\end{figure}

Regarding the steady state at large dissipation and $r=1$, in Figure~\ref{fig:SM_r1_purity} we show the evolution of the variance of steady-state eigenvalues as a function of $g_\mathrm{eff}$. For $r=1$ the condition  $\nu_{C''}=\nu_{D''}$ is no longer verified, instead $\nu_{C''}=-2$. Using the values of $\nu_{P''}$ and $\kappa_{\rho_0}^<$ obtained from Figure~\ref{fig:SM_r1_purity}~$(a)$, and of $\nu_{D''}$ and $\kappa_{\rho_0}^>$ obtained from Figure~\ref{fig:SM_r1_purity}~$(c)$, (\ref{eq:exponents_constraint}) gives the value $\lambda_{\rho_0}=8/5$. Note that in this case, both boundaries of regime C$''$ scale with $N$ and the scaling function does not have a single power-law behaviour throughout the intermediate regime. 
However, the power-law $g_\mathrm{eff}^{\lambda_{\rho_0}}$ describes accurately the behaviour of $\sigma_{\rho_0}^2$ both when entering regime C$''$ from regime P$''$ and when entering regime C$''$ from regime D$''$, see the dashed lines in Figure~\ref{fig:SM_r1_purity} $(b)$. The two power-laws are shifted with respect to each other and are linked by a crossover regime in the center of regime C$''$.

Finally, we have $\sigma^2_{\rho_0}\propto1/N$ for strong dissipation, which gives a purity $\mathcal{P}_0\propto N^0$ in the large-$N$ limit. Thus, contrarily to the $r>1$ case, the steady-state purity is not proportional to that of the maximally mixed state but remains finite for large $N$.

\begin{ack}
We gratefully acknowledge valuable discussions with Tankut Can. LS acknowledges support by FCT through PhD Scholarship SFRH/BD/147477/2019. PR acknowledges support by FCT through the Investigador FCT contract IF/00347/2014 and Grant No.\ UID/CTM/04540/2019. TP acknowledges ERC Advanced Grant 694544-OMNES and ARRS research program P1-0402.
\end{ack}

\section*{References}

\bibliographystyle{iopart-num}
\bibliography{RL}

\end{document}